\documentclass[twocolumn,showpacs,amsmath,amssymb,floatfix]{revtex4}

\usepackage{graphicx}
\begin{document}

\title{Temperature dependence of spectral functions for the one-dimensional
  Hubbard model: comparison with experiments}
\author{A. Abendschein$^{(1),(2)}$ and F. F. Assaad$^{(1)}$}

\affiliation{$^{(1)}$ Institut f\"ur theoretische Physik und Astrophysik,
Universit\"at W\"urzburg, Am Hubland D-97074 W\"urzburg, Germany \\
   $^{(2)}$ Laboratoire de Physique Th\'eorique, IRSAMC, Universit\'e Paul Sabatier,
     118, route de Narbonne,31062  Toulouse, France} 

\begin{abstract}
We study the temperature dependence of the single particle spectral function 
as well as of the  dynamical spin and charge structure factors for the
one-dimensional Hubbard model using the finite temperature auxiliary field quantum 
Monte Carlo algorithm. The parameters of our simulations are chosen so to at best 
describe the low temperature photoemission spectra of the organic conductor TTF-TCNQ. 
Defining a magnetic energy scale, $T_J$,  which  marks the onset of short ranged 
$2k_f$ magnetic  fluctuations,  we conclude that for temperatures $T < T_J$ 
the ground state features of the single particle spectral function are apparent in the finite 
temperature data. Above $T_J$ spectral weight transfer over a scale set by the hopping 
$t$ is observed.  In contrast,  photoemission data points to a lower energy scale below which 
spectral weight transfer occurs. Discrepancies between Hubbard model calculations 
and experiments are discussed. 
\end{abstract}

\pacs{71.27.+a, 71.10.-w, 71.10.Fd}

\maketitle

\section{Introduction}
\label{introduction.sec}
	The hallmark of Luttinger liquids lies in spin-charge separation; an 
electron fractionalizes into a spinon carrying the spin degrees of freedom and a 
holon carrying the charge. Detecting spin-charge separation relies on the 
study of the single particle spectral function. From the theoretical point of view,
the ground state low energy properties of the single particle spectral 
function can be obtained from  bosonization \cite{Giamarchi}.  This approach   
yields two branch cuts, corresponding to the spinon and the holon, 
dispersing linearly from the Fermi wave vector 
with spin and charge velocities.  
Beyond this low energy limit, exact calculations of the spectral function have 
been carried out for the  infinite $U$ Hubbard model \cite{Penc96}. Taking 
into account that in this limit the spin velocity vanishes, the results stand 
in agreement with the low energy bosonization picture. Furthermore the calculations 
reveal higher energy features such as a holon shadow band. Hence, distinct signatures
of  Luttinger liquids may be found in a wide energy range thus facilitating  detection 
in photoemission experiments. Beyond the infinite $U$ limit, numerical simulations 
such as  dynamical DMRG (DDRMG) \cite{Jeckelmann04} or Quantum Monte Carlo \cite{Preuss94} can 
be used to  investigate the zero temperature properties of the spectral function.    
In particular, $T=0$ DDMRG results  for  the Hubbard model 
have been compared successfully  with low temperature, $T=60K$, photoemission experiments  
on the organic one-dimensional conductor TTF-TCNQ \cite{Claessen03}. 
In the temperature range $60 K < T < 260 K$ experiments 
point towards substantial  spectral weight transfer.
Keeping  the model parameters which reproduce the low temperature 
data, our aim is  to understand if the experimentally observed 
temperature behavior of the spectral function can be reproduced by finite temperature 
model calculations.  

The Hubbard model we consider reads:
\begin{eqnarray}
H &=& -t \sum_{\langle i,j\rangle, \sigma}
(c^{\dagger}_{i\sigma}c_{j\sigma} + h. c.) + \nonumber \\
&& \hphantom{lll} +    U \sum_{i} n_{i\uparrow} n_{i\downarrow} 
- \mu \sum_{i} (n_{i\uparrow} + n_{i\downarrow})
\label{h_hubbard}
\end{eqnarray}
where $t$  it the hopping amplitude,  $U$ the Coulomb repulsion, $\mu $ the chemical potential, 
and the first sum  runs over nearest neighbors. 
$c^{\dagger}_{i\sigma}$ ($c_{i\sigma}$)  creates
(annihilates) an electron in the Wanier state centered around lattice site $i$ and 
with $z$-component of spin $\sigma = \uparrow, \downarrow$.
Comparison  between DDMRG results and experiments point to a parameter set  
$ U/t = 4.9  $, $ t = 0.4 eV$  and  $n= 0.59 $  for an adequate description of the 
TCNQ chain.  Throughout this article we will keep those parameters fixed  and vary the 
temperature. 
The organization and main results of the paper are as follows. In section \ref{montecarlo.sec}
we briefly present the finite temperature auxiliary field quantum Monte Carlo (QMC) 
method and the
maximum entropy  method we have used to analytically continue the 
imaginary time QMC data. Section \ref{results.sec} is dedicated to the results.  To map 
out the scales involved in the problem, we first consider the temperature dependence of 
the spin and charge susceptibilities as well as of the spin and charge dynamical 
 structure factors. 
This allows us to define a magnetic crossover energy scale, $T_J$, below which the magnetic 
$2k_f$ correlation length increases substantially  as a function of decreasing temperature. 
In section \ref{sp_excitation.sec} we analyze  the temperature dependence of the single 
particle  spectral  function and arrive to the conclusion that $T_J$ is the only low energy 
scale at hand in  the problem. That is,  for temperatures below $T_J$ the overall features of 
the zero  temperature spectral function are well reproduced. Above $T_J$ we observe spectral 
weight transfer over energy scale set by $t$.  With $t = 0.4eV$ our 
estimated value of $T_J$ is $T_J \simeq 400 K$. In the conclusion 
(Sec. \ref{conclusion.sec}),
we discuss possible scenarios to understand the discrepancy between experimental data 
and model  calculations. 

\begin{figure}
\includegraphics[width=.4\textwidth]{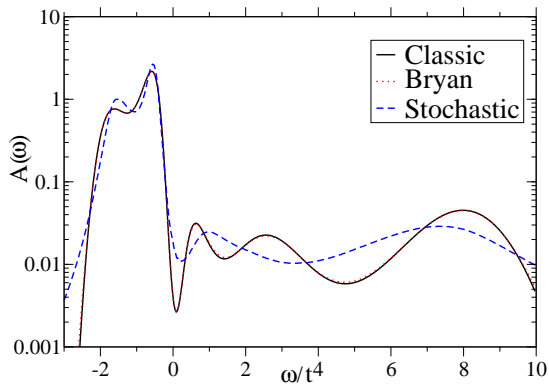}  
\caption{Comparison of different Maximum Entropy methods for the spectral function at $k = 0$, 
of the one-dimensional Hubbard model at $U/t = 4.9$, $n = 0.59$ and $\beta t = 7$.  Note the 
semi-log scale. } 
\label{Aom_comp.fig}
\end{figure}

\begin{figure}
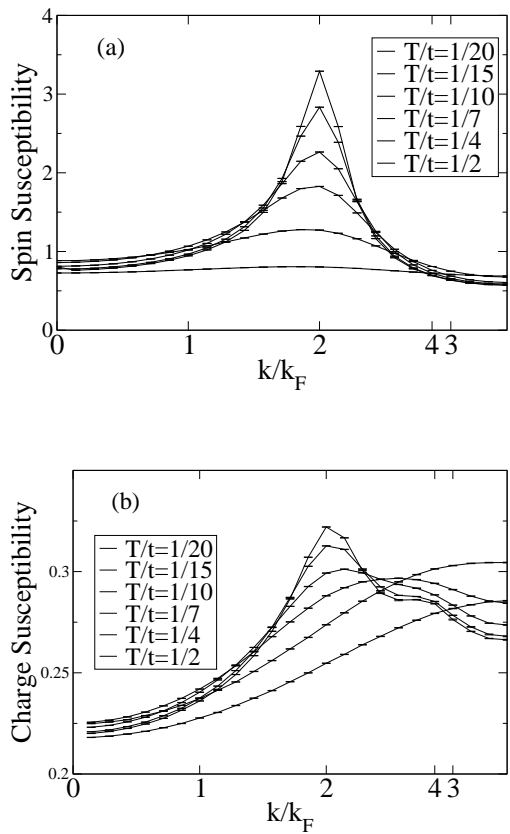

\includegraphics[width=.37\textwidth]{Figs/spin_suscep.eps}   \\
\vspace*{1.0cm}
\includegraphics[width=.37\textwidth]{Figs/charge_suscep.eps} \\ 
\caption{Spin  and charge static susceptibilities, $\chi_{\stackrel{s}{c}}(k,\omega=0)$,  as a function of wave vector. At $2k_f$ the 
different lines from bottom to top correspond to the temperatures  
$\beta t = 2, 4, 7, 10, 15, 20$} 
\label{Susceptibilities.fig}
\end{figure}

\section{Quantum Monte Carlo Algorithm}
\label{montecarlo.sec}
We have used a generic implementation of the finite temperature grand canonical 
auxiliary field algorithm  \cite{Assaad02}  to  compute  imaginary time displaced 
 Green functions, $ G(\vec{k}, \tau)  = 
\langle c_{\vec{k} } (\tau) c^{\dagger}_{\vec{k}}(0) \rangle $,  as a function of 
temperature for the one-dimensional Hubbard model
on a $48$-site chain with periodic boundary conditions.  
The spectral function is extracted  from the imaginary time data  by inverting  the 
equation:
\begin{eqnarray}
 & & \langle c_{\vec{k}} (\tau) c^{\dagger}_{\vec{k}} (0) \rangle =  \int {\rm d} \omega
    K(\tau,\omega)   A(\vec{k}, \omega)   \; \;  {\rm with} \nonumber \\
 & & K(\tau,\omega)  = \frac{1}{\pi}  \frac{e^{-\tau \omega}  }
        { 1  +  e^{- \beta \omega}}.
\end{eqnarray}
Since this inversion is numerically ill defined we have used the Maximum Entropy method and 
favored a recently proposed  stochastic version \cite{Beach04a}.  The 
stochastic formulation  has the appealing property that it 
formally contains the generic  Maximum Entropy method (MEM)  \cite{Jarrell96,Linden95} at the 
{\it mean field} level. In the generic MEM, there is no free parameter.
In particular,  $\alpha$ which determines how much information is  taken from the default 
model is determined self-consistently. In contrast, in the stochastic approach, 
there is no sharp  way of determining  $\alpha$ and we have used the criterion proposed by 
K. Beach \cite{Beach04a}.  In Fig. \ref{Aom_comp.fig}  we compare the different 
Maximum Entropy methods for the one-dimensional Hubbard model at  $U/t = 4.9$, $n = 0.59$ 
and $\beta t = 7$.  The Brayn and Classic Maximum entropy formulations \cite{Jarrell96}  yield
identical spectral function.  The two peak structure at $\omega/t < 0 $ corresponding to the 
holon and spinon branches, is sharper in the stochastic approach. 
At $\omega/t > 0$ the  stochastic  spectrum shows less features than the Classic 
Maximum Entropy method. It is know that the Classic Maximum Entropy has difficulties in 
reproducing flat spectra and generates
a curve oscillating smoothly around the correct value. This problem seems to be alleviated  
by the stochastic approach. 
Hence, our overall opinion is that the stochastic approach does better at reproducing 
sharp features as well as flat regions in the spectra. 
Finally we note that we have always taken the covariance matrix into account. 

\section{Results}
\label{results.sec}
We compute the single
particle spectral function as well as the dynamical spin and charge structure 
factors as a function of temperature. We choose a parameter set which at best describes the 
low temperature properties of the photoemission spectra of TTF-TCNQ \cite{Claessen03}. 
That is  for the TCNQ band, $U/t = 4.9$,  $t = 0.4$ eV  and a filling fraction $n = 0.59$. 
We cover the following range of inverse temperatures
$\beta t = 2, 4, 7, 10, 15$ and $\beta t = 20$. Below, we will first discuss two 
particle properties so as to pin down scales and then consider the temperature 
behavior of the single particle spectral function.  
 
\begin{figure}
\begin{center}
\begin{tabular}{cc}
\resizebox{85mm}{!}{\includegraphics{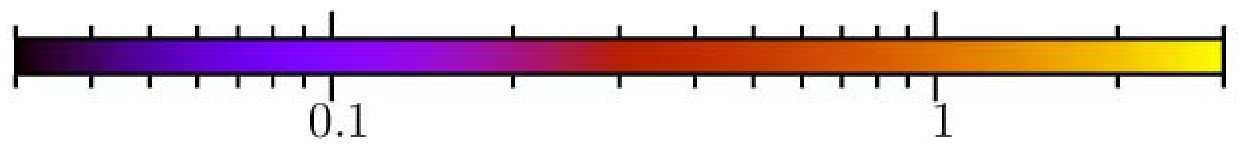}} \\
\resizebox{85mm}{!}{\includegraphics{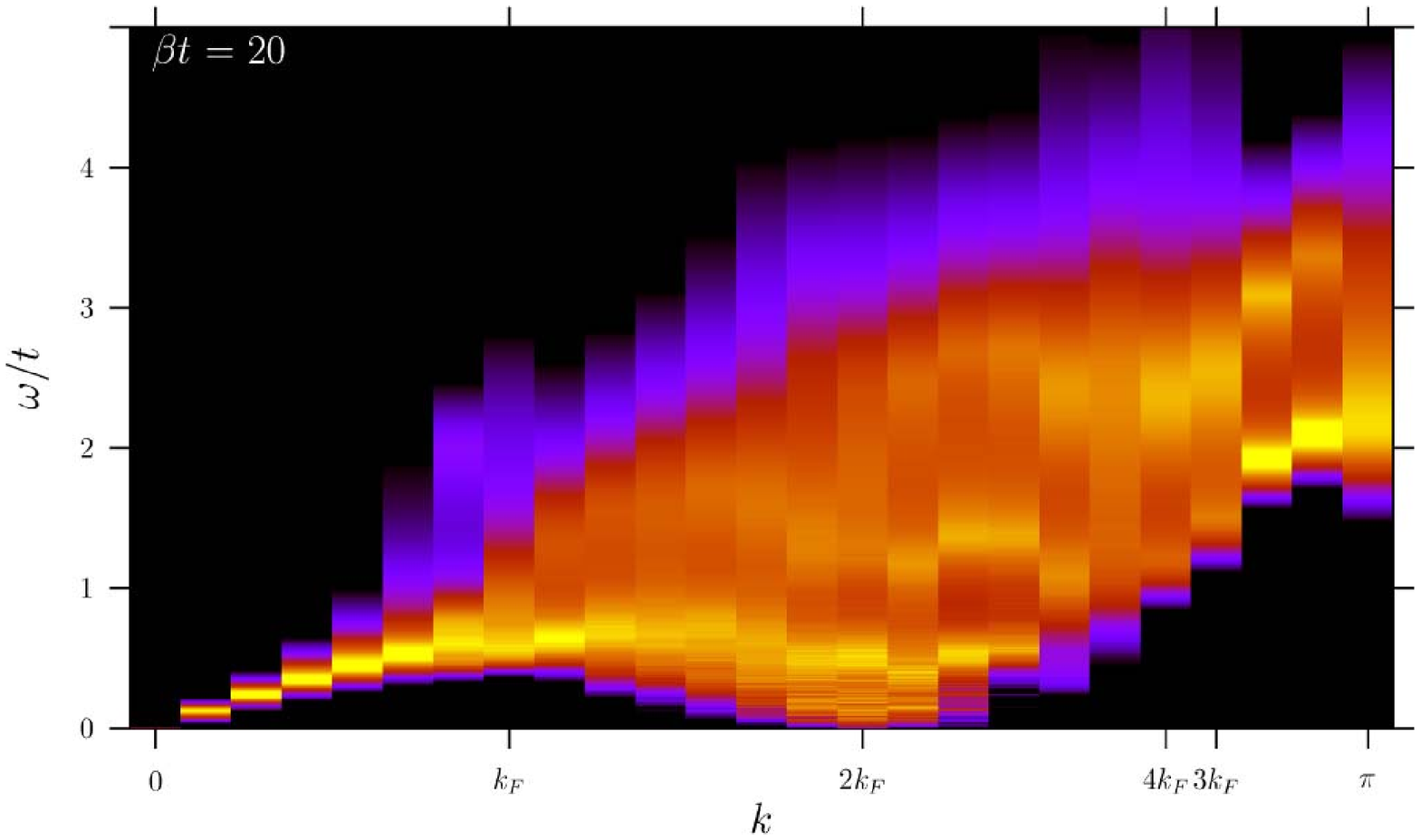}} \\ 
\resizebox{85mm}{!}{\includegraphics{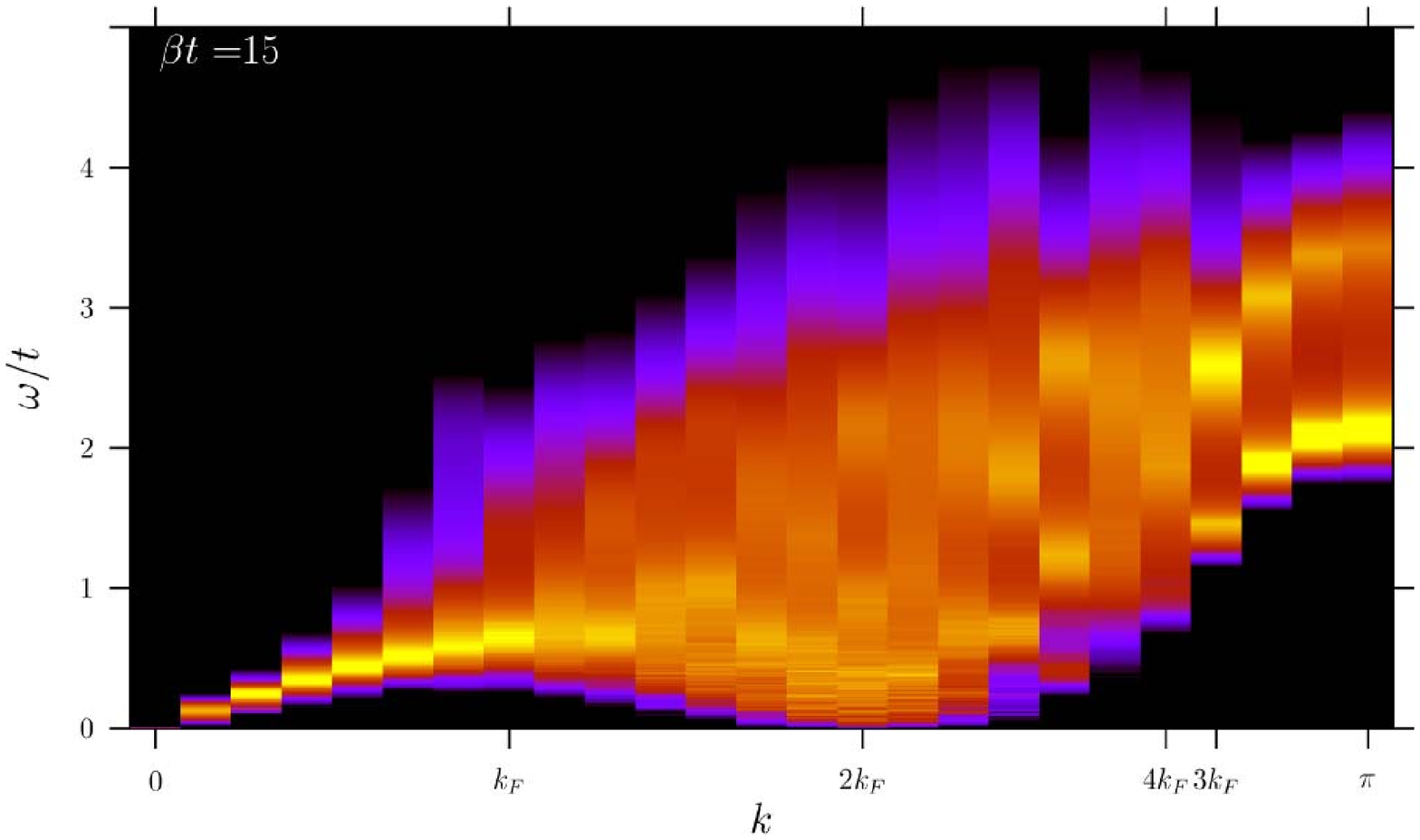}} \\
\resizebox{85mm}{!}{\includegraphics{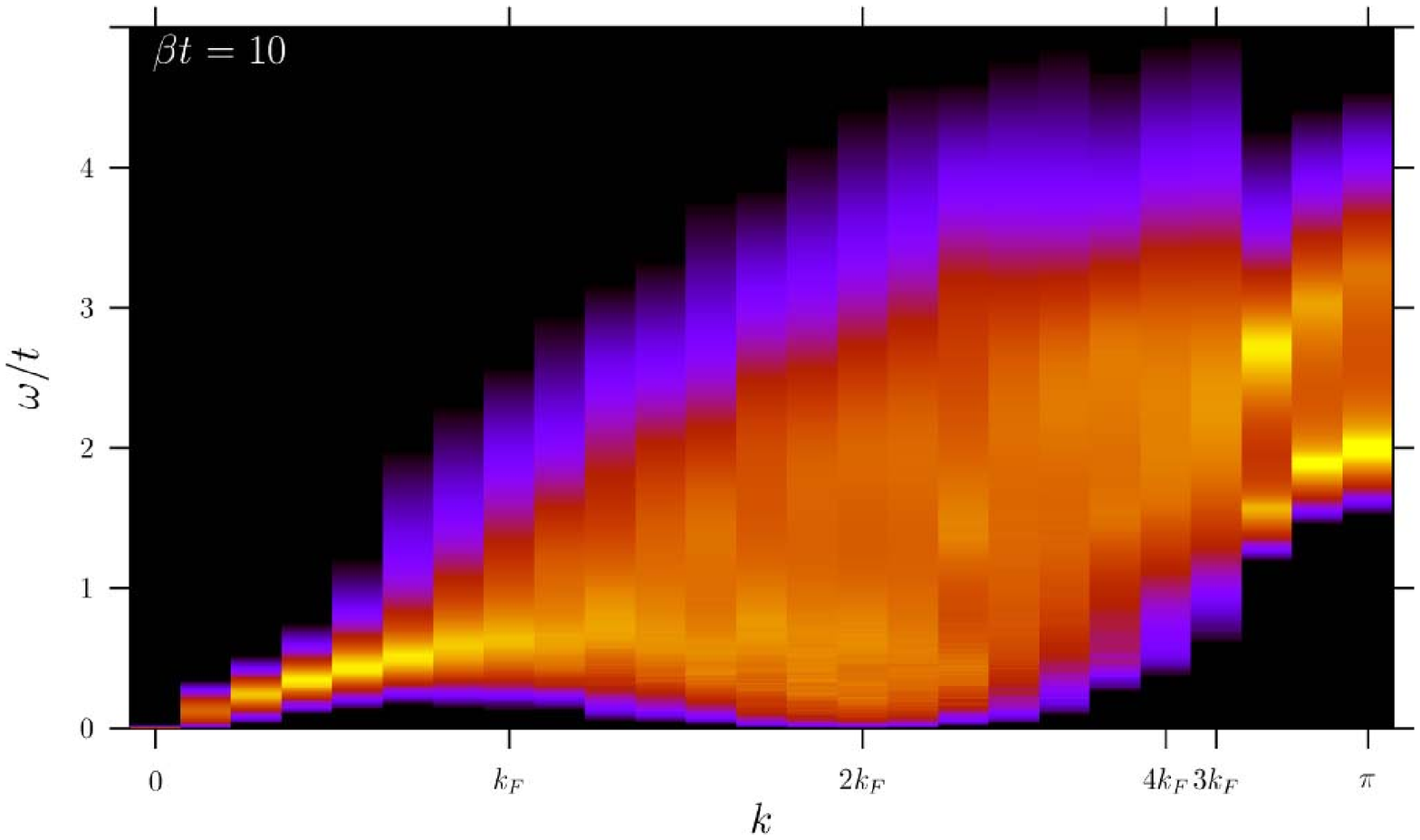}} \\
\resizebox{85mm}{!}{\includegraphics{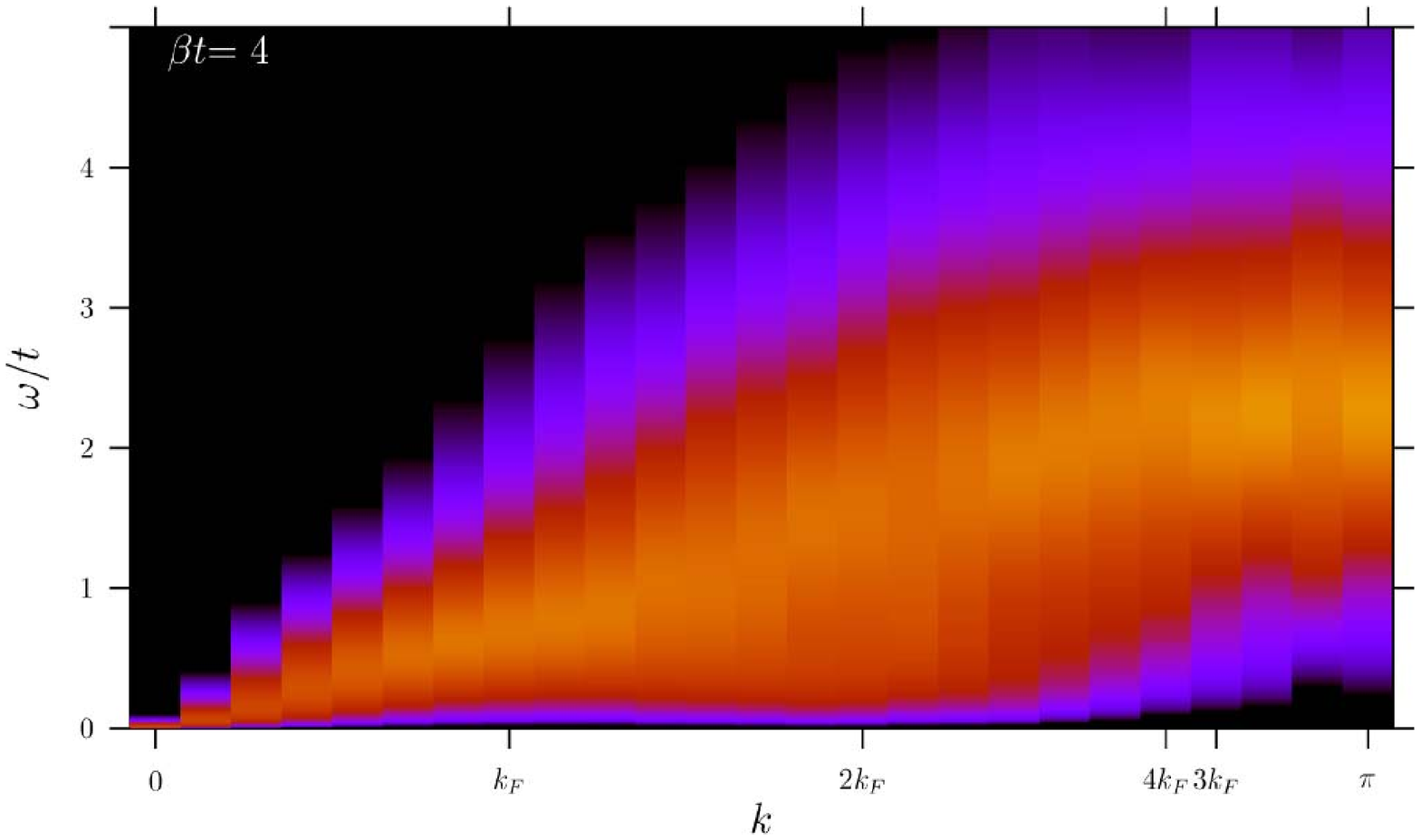}} 
\end{tabular}
\caption{Dynamical spin-spin correlations as a function of temperature on a logarithmic
intensity scale.  The inverse temperatures from 
top to bottom read:  $\beta  t=20,15,10, 4$.} 
\label{spin_log_temp_1.fig}
\end{center}
\end{figure}

\begin{figure}
\begin{center}
\begin{tabular}{cc}
\resizebox{85mm}{!}{\includegraphics{Figs/Spin_Charge_log_rgb_scale.eps}} \\
\resizebox{85mm}{!}{\includegraphics{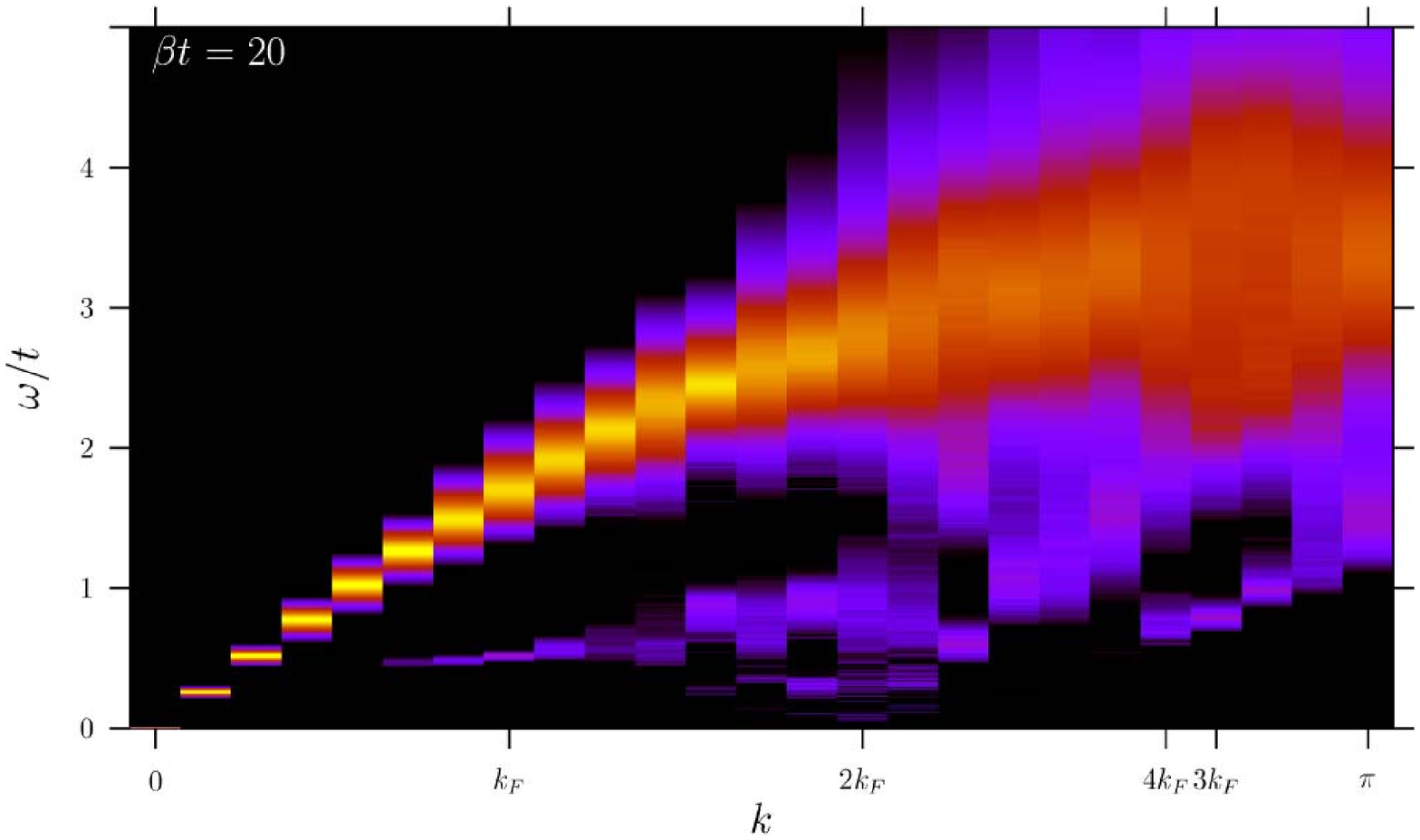}} \\ 
\resizebox{85mm}{!}{\includegraphics{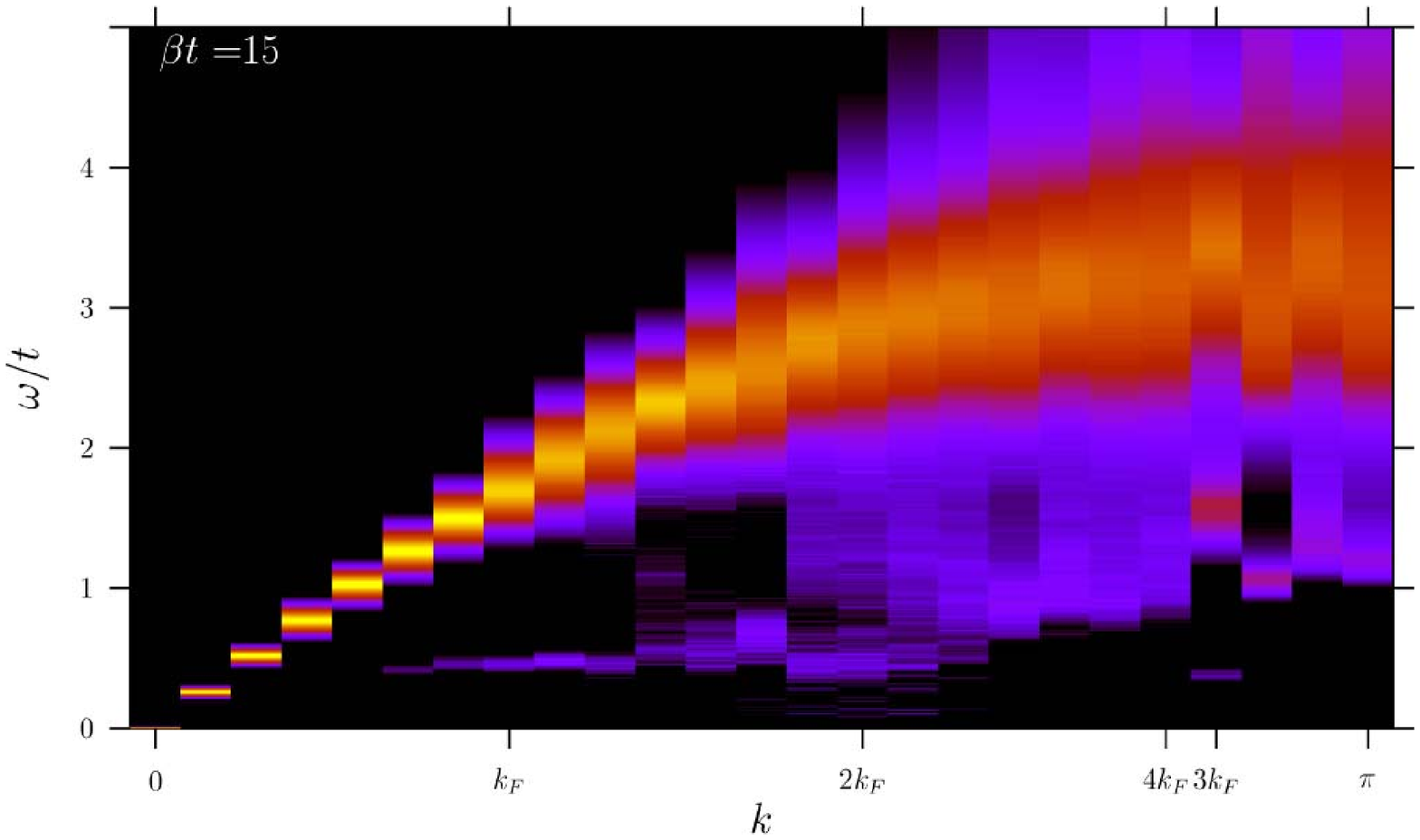}} \\
\resizebox{85mm}{!}{\includegraphics{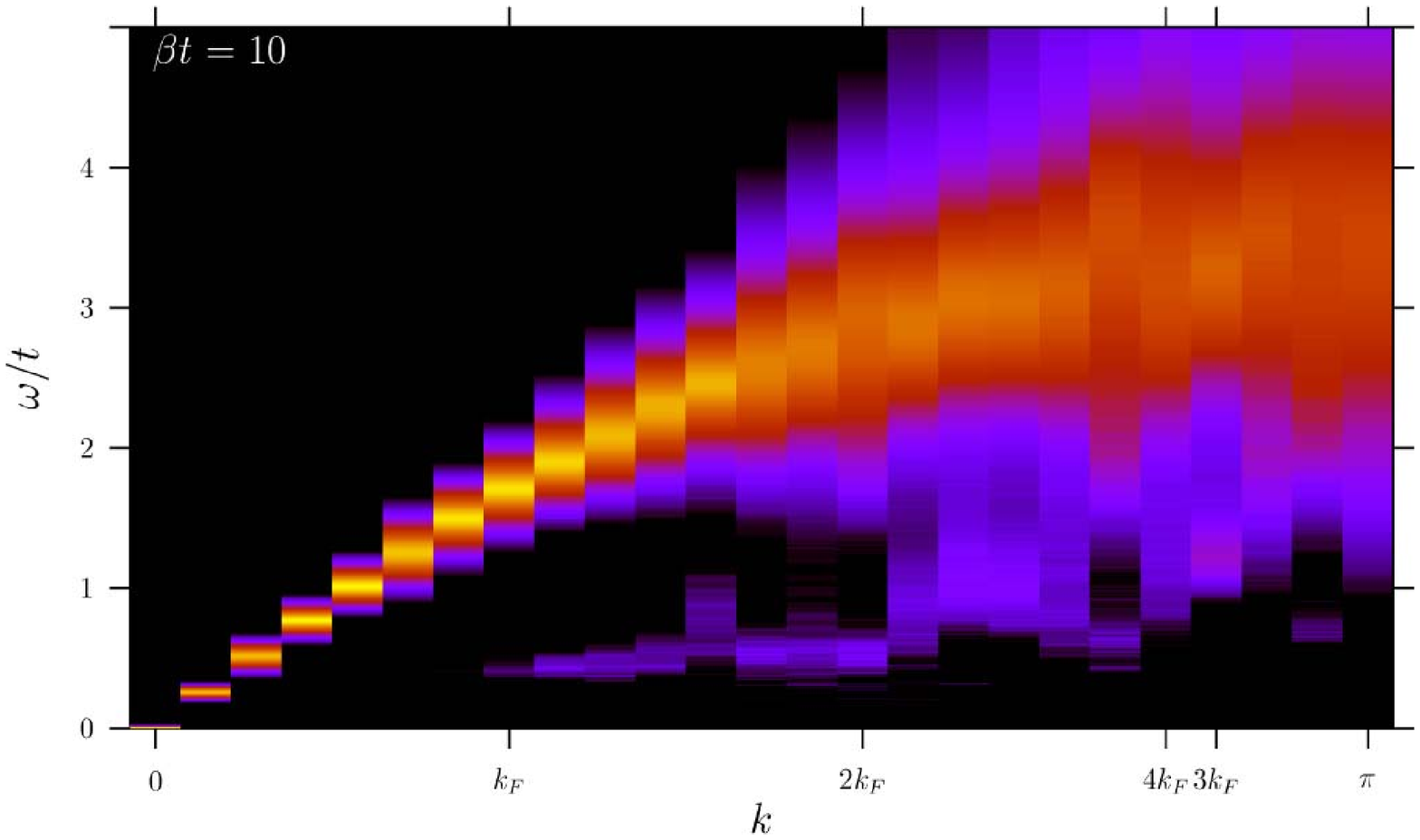}} \\
\resizebox{85mm}{!}{\includegraphics{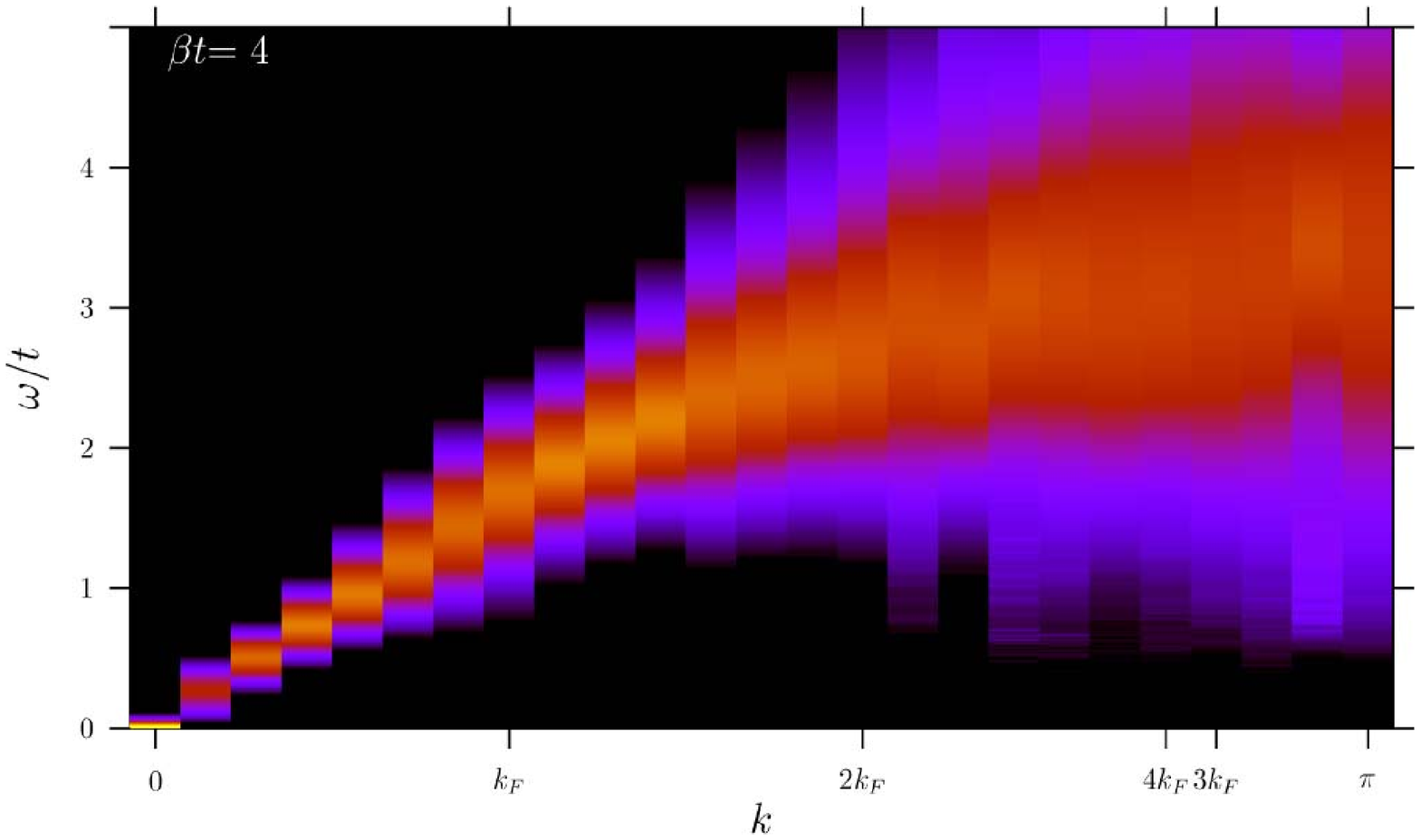}}
\end{tabular}
\caption{Dynamical charge-charge correlations as a function of temperature on a logarithmic
intensity scale.  The inverse temperatures from 
top to bottom read:  $\beta  t=20,15,10, 4$.} 
\label{charge_log_temp_1.fig}
\end{center}
\end{figure}

\subsection{Dynamical spin and charge correlation functions.}
\label{spch_excitations.sec}
To investigate the spin and charge dynamics we consider the dynamical susceptibility
\begin{equation}
	\chi_{\stackrel{c}{s}}(k,\omega) =  - i \int_{0}^{\infty} {\rm d} t e^{i \omega t} 
 \langle \left[ O_{\stackrel{c}{s}} (k, t), O_{\stackrel{c}{s}}(-k,0)  \right] 
 \rangle
\end{equation}
where $O_{\stackrel{c}{s}}(\vec{k}) = 
\frac{1}{\sqrt{N}} \sum_{\vec{r}} e^{i \vec{k} \cdot \vec{r} }  
\left(  n_{\vec{r},\uparrow} \pm n_{\vec{r},\downarrow} \right) $.
Fig. \ref{Susceptibilities.fig} plots the static spin and charge susceptibilities, 
$\chi_{\stackrel{c}{s}} (\vec{k}, \omega = 0 ) $ as a function of temperature. 
Those quantities measure correlation lengths and 
allow us to identify crossover energy scales below which spin and charge fluctuations  
grow as a function of decreasing temperature. Since one dimensional systems are 
critical at $T=0$ both $2 k_f$ spin and charge static susceptibilities 
diverge at $T=0$.       
As apparent from Fig. \ref{Susceptibilities.fig}    
the crossover scale marking the growth of $2k_f$ spin fluctuations is given 
by $T_J \simeq t/10 $. A similar energy scale for the $2k_f$ charge fluctuations can be 
read off Fig. \ref{Susceptibilities.fig}. However the magnitude of the signal, and hence 
the amplitude of  the $2 k_f$ charge modulation is substantially smaller than in the spin 
sector.   
The Luttinger liquid parameter, $K_\rho$ of the Hubbard model is bounded by $ 1/2 < K_\rho < 1$ 
\cite{Schulz90}. Hence $4k_f$ charge-charge correlations which decay as 
$ r^{-4 K_\rho} $ are very much suppressed in comparison to $2 k_f$ charge fluctuations 
which decay as $ r^{-1 - K_\rho} $. This stands in accordance with the data of Fig. 
\ref{Susceptibilities.fig} and no divergence in the $4k_f$ charge susceptibility is expected. 

Having pinned down energy  
scales  we now consider the dynamical spin and charge structure factors:
\begin{equation}
	S_{\stackrel{c}{s}}(\vec{k},\omega) 
        = \frac{1}{1-e^{-\beta \omega} } {\rm Im} \chi_{\stackrel{c}{s}}(\vec{k}, \omega).
\end{equation}
Fig. \ref{spin_log_temp_1.fig} plots the dynamical spin structure factor  as a function of 
temperature. As apparent below the crossover scale $T_J$ the two spinon continuum of 
excitations with gapless excitations at $2k_F$ is clearly visible.  Furthermore, below $T_J$, a
well defined spin velocity can be read off the data yielding 
$v_s/t \simeq 1$. This result
compares favorably with the zero temperature results of Ref.  \cite{Schulz95}. Hence, in the 
spin sector $T_J$ marks the temperature scale below which the the  overall features of the 
zero temperature dynamical spin structure factor become apparent. 

The dynamical charge structure factor is plotted as a function of temperature  in 
Fig. \ref{charge_log_temp_1.fig}. Again, below $T_J$, one can read off the charge velocity, 
$v_c \simeq 1.9 $ which favorably compares with  the zero temperature data of \cite{Schulz95}. 
Given the small amplitude of the $2k_f$ charge fluctuations, we are unable to reliably pin down 
the expected gapless excitations at $2 k_f$ as well as at $4 k_f $.

\subsection{Single particle excitation spectrum}
\label{sp_excitation.sec}
\begin{figure}
\begin{center}
\begin{tabular}{cc}
\resizebox{85mm}{!}{\includegraphics{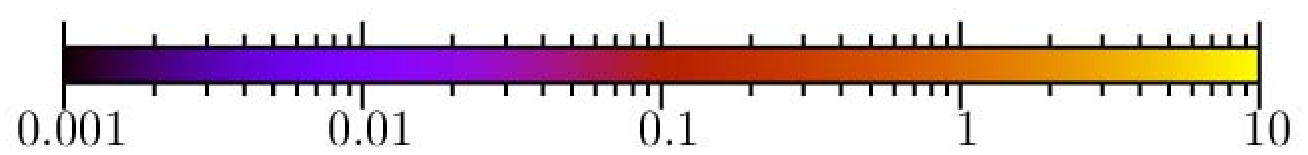}} \\
\resizebox{85mm}{!}{\includegraphics{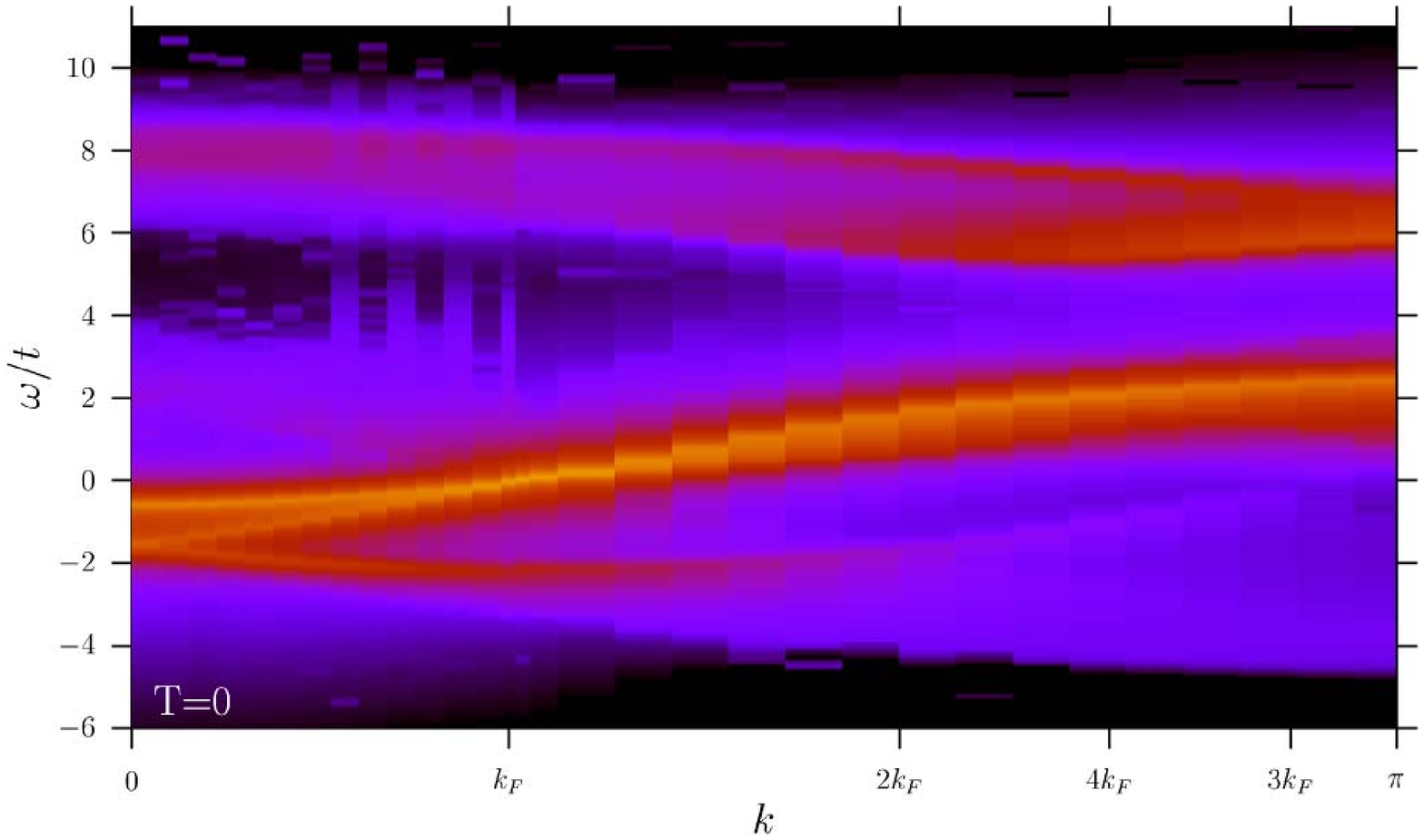}}
\end{tabular}
\caption{Single particle excitation spectrum for $T=0$ shown as gray scale
  plot with a  logarithmic intensity scale. These data stem from the 
DDMRG calculations of H. Benthien et al. \cite{Jeckelmann04} }
\label{green_temp_1.fig}
\end{center}
\end{figure}

Our major interest here is to study the temperature dependence of  the single 
particle spectral function and compare to the experiments of Ref. \cite{Claessen03}.
At the lowest temperatures considered in Ref. \cite{Claessen03}, the photoemission 
results compare favorably with $T=0$ DDMRG calculations of Ref. \cite{Jeckelmann04}. 
In the photoemission spectra one can identify a spinon branch, a
holon branch as well a holon shadow band. 
Those features compare well with the zero temperature DDMRG data shown in 
Fig. \ref{green_temp_1.fig}. Let us concentrate on $\omega/t < 0$ relevant for  comparison
with photoemission. In the vicinity of the Fermi wave vector and  at low
energies  one clearly observes two features (branch-cuts) dispersing linearly 
with velocities  $v_s$ (spinon) and $v_c$ (holon). Those velocities stand in good 
agreement with those determined by our analysis of the spin and charge dynamical 
structure factors.  
Furthermore, and at low energies one can identify a feature at  $3k_f$ which merges   
at $k=0$ with the holon branch.  Following Ref. \cite{Penc96}  one notes that this feature has the
same dispersion relation as the holon branch but shifted by $2 k_f$. Hence the interpretation 
of a shadow holon branch which stems from a holon scattering off a $2k_f$ spin excitation. 
This is very reminiscent of shadow bands in spin-density wave approaches of antiferromagnetic 
Mott insulators.

\begin{figure}
\begin{center}
\begin{tabular}{cc}
\resizebox{85mm}{!}{\includegraphics{Figs/Green_rgb_scale.eps}} \\
\resizebox{85mm}{!}{\includegraphics{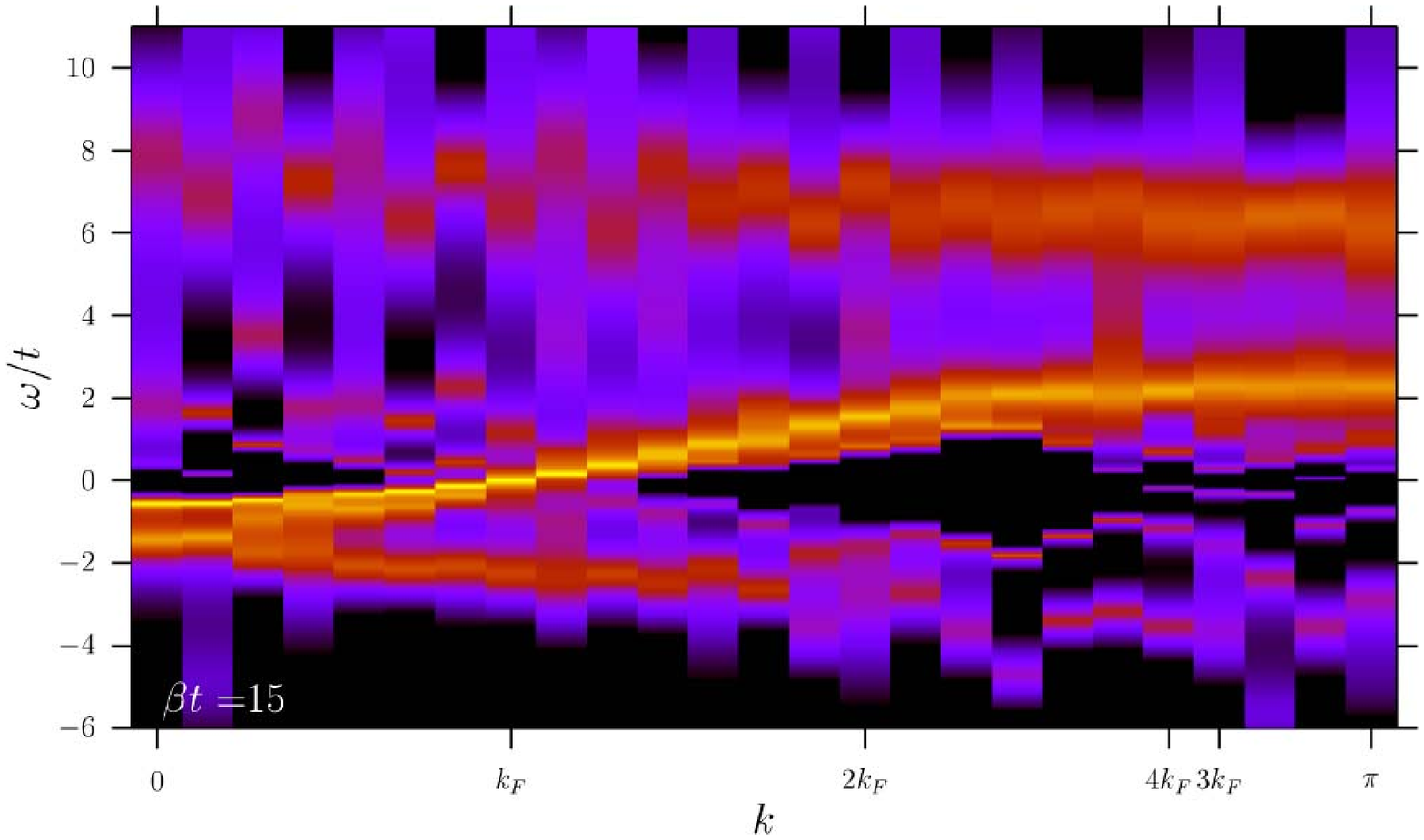}} \\
\resizebox{85mm}{!}{\includegraphics{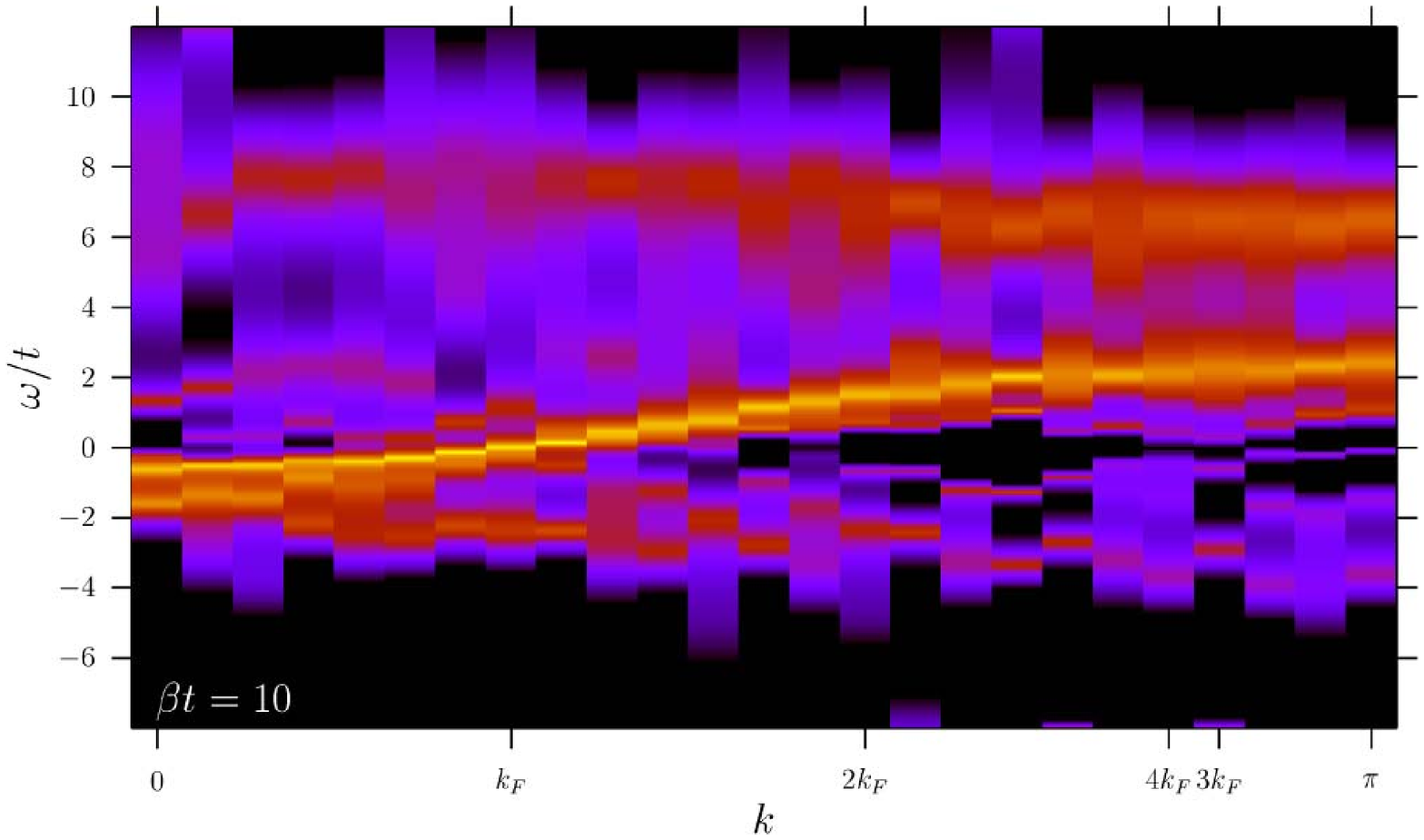}} \\
\resizebox{85mm}{!}{\includegraphics{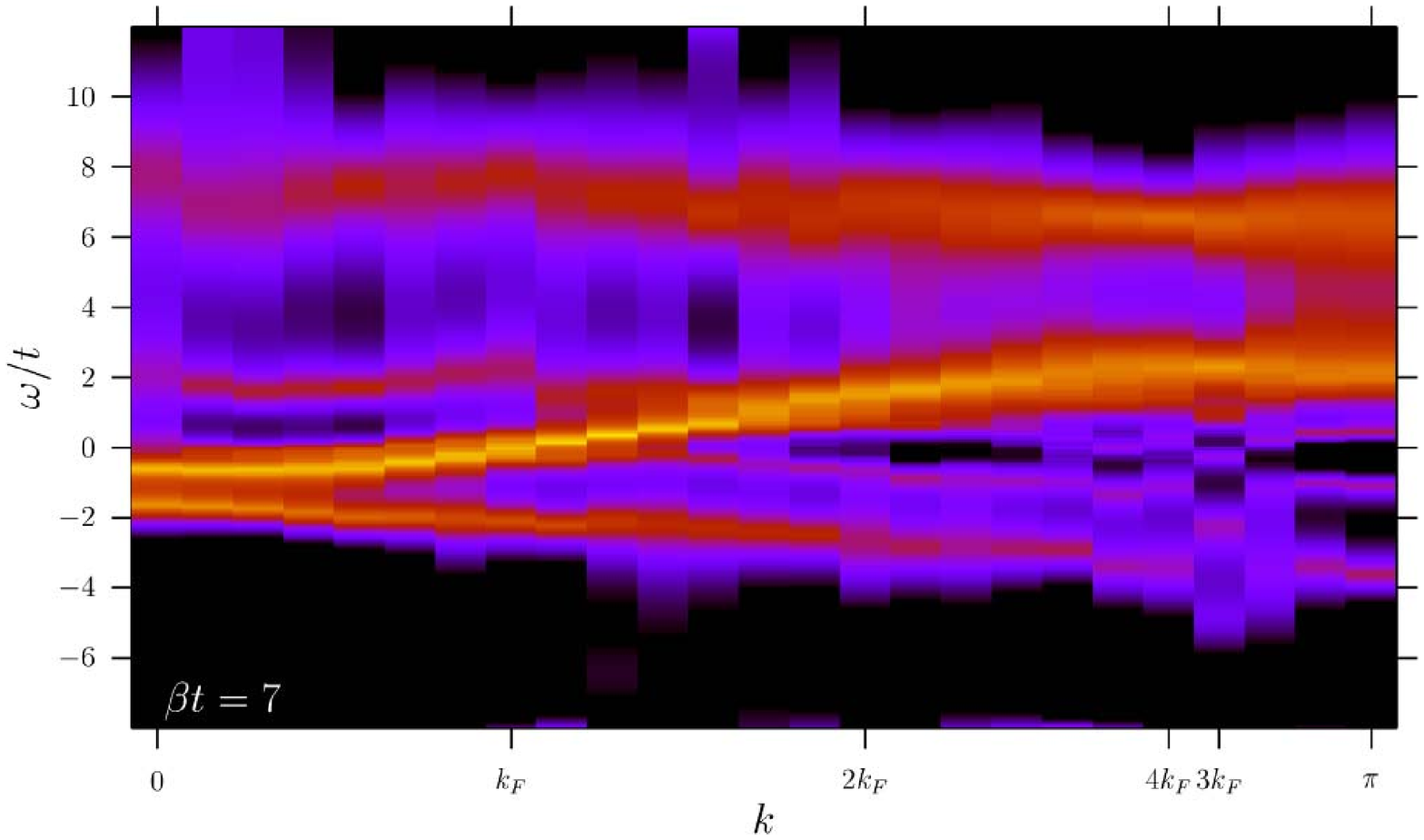}}  \\
\resizebox{85mm}{!}{\includegraphics{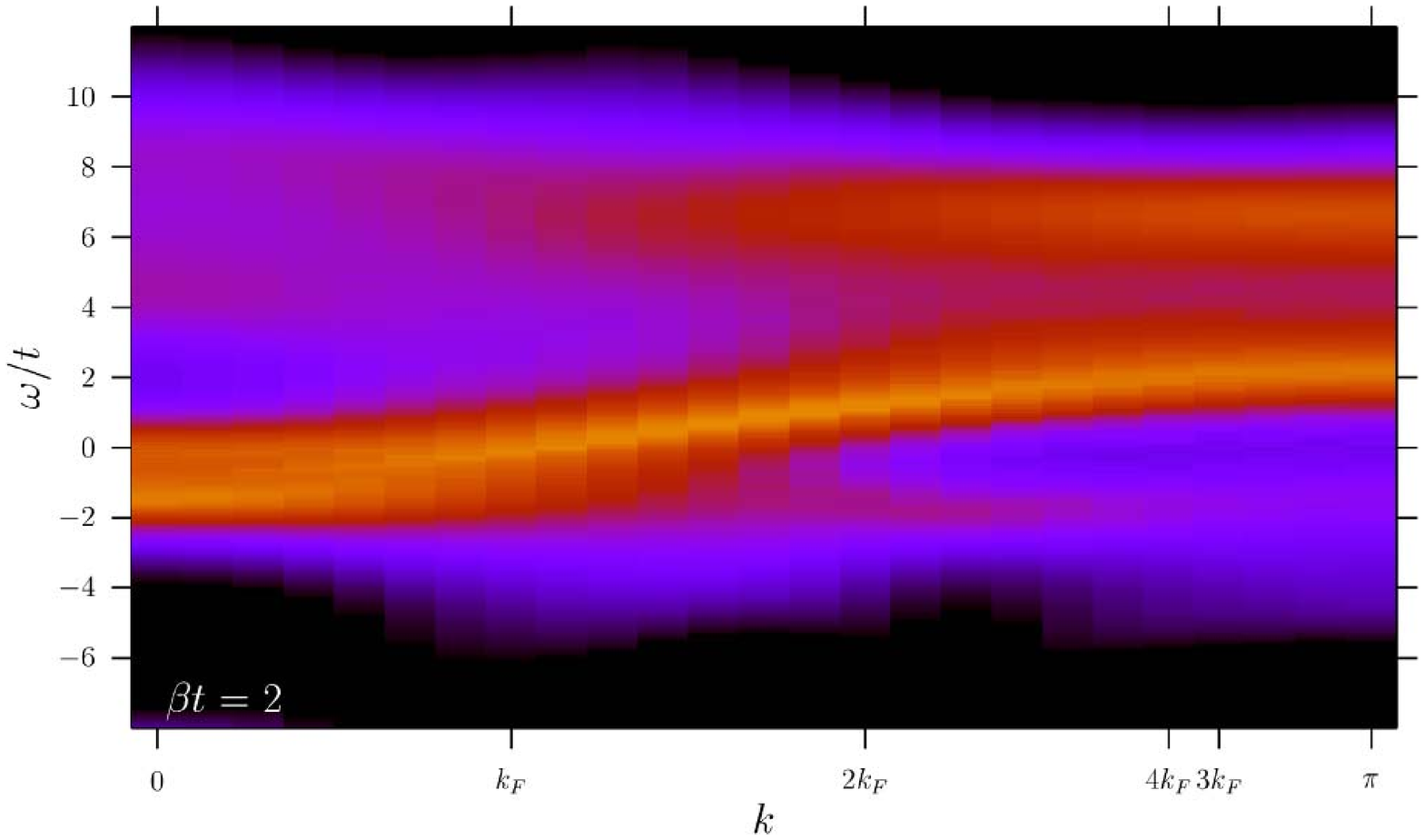}} 
\end{tabular}
\caption{Single particle excitation spectrum as a function of temperature on 
a logarithmic intensity plot.  From top to bottom: $\beta t=15, 10, 7, 2$. }
\label{green_temp_2.fig}
\end{center}
\end{figure}

The finite temperature spectra we have  obtained with the QMC are 
presented in Fig. \ref{green_temp_2.fig}.  The question we wish to address is 
at which temperature scale do the features of the $T=0$ data become apparent. 
One can observe a clear  spinon branch for the different temperatures
except at $\beta t = 2$ where spectrum is  washed out. The
holon shadow band can also be identified at least in the region of wave
vectors where $0 \leq k \leq k_F$. As can be seen in the DDMRG $T=0$
spectrum, the intensity of the holon shadow band  rapidly decreases for
larger wave vectors making it very hard to retrace this feature in our
spectra. Another difficulty arises regarding the holon branch in our QMC
spectra. To analyze our data we use the stochastic MEM,  with its well-known difficulties
to resolve two peaks close in energy. Nevertheless, at our lowest 
presented temperature $\beta t = 15$, a holon branch can be identified. 
Hence, the finite temperature results stand in agreement with the statement that 
below the spin scale $T_J$, the gross  features of the zero temperature results
are apparent.

\begin{figure}
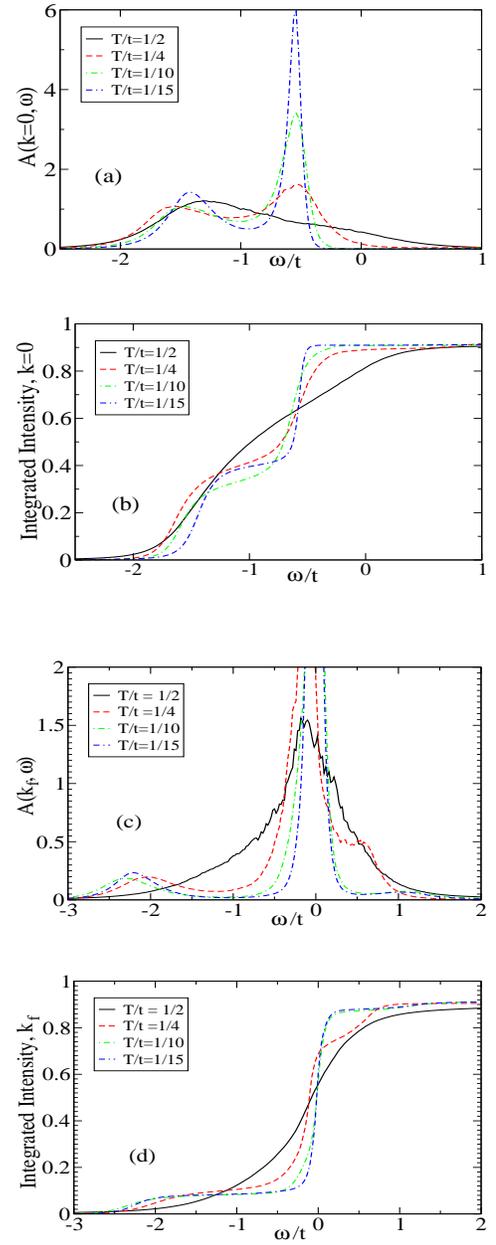

\begin{center}
\includegraphics[width=.35\textwidth,height=0.15\textheight]{Figs/k_0_tmp.eps}  \\
\vspace*{0.6cm}
\includegraphics[width=.35\textwidth,height=0.15\textheight]{Figs/k_0_int.eps}  \\
\vspace*{1.0cm}
\includegraphics[width=.35\textwidth,height=0.15\textheight]{Figs/k_f_tmp.eps}  \\
\vspace*{0.6cm}
\includegraphics[width=.35\textwidth,height=0.15\textheight]{Figs/k_f_int.eps} 
\caption{Temperature dependence of spectral functions at (a) $k=0$   and (c) $k=k_f$.
Integrated spectral weight for (b) $k=0$ and (c) $k=k_f$.  }
\label{green_spect_temp.fig}
\end{center}
\end{figure}

The  ARPES measurements of Ref. \cite{Claessen03} point towards
spectral weight transfer in the temperature range $60 K < T < 260 K$. At the 
Fermi wave vector and as a function of decreasing temperature spectral weight 
is transfered  from higher ($ \simeq 0.7 eV$) to lower ($\simeq 0.1 eV$) excitation energies. 
The numerical simulations do indeed show spectral weight transfer however at a temperature  
scale $  T > T_J$.  Figs. \ref{green_spect_temp.fig}  plots $A(k,\omega)$ as a function of temperature for $ k = 0 $ and $k = k_f$.  Upon inspection of the data, one observes that at high 
temperatures ($\beta t = 2$), the spectral weight is  dominantly located at frequency of 
the holon,  $\omega/t \simeq -1.5 $ for $k  = 0$ and $\omega \simeq 0$ at $k = k_f$. As 
temperature  is lowered thereby  generating short ranged $2k_f$ spin fluctuations, spectral 
weight is shifted  over an energy scale set by $t$ to form the  spin related features. At 
$k=0$, this corresponds to the spinon at $\omega/t \simeq -0.5$ and at $k = k_f$ to the 
holon-shadow band at $\omega/t \simeq -2$.
For $T < T_J$   and within the limitations of the stochastic analytical 
continuation, the data shows no further shift of spectral weight.

\section{Conclusion}
\label{conclusion.sec}
We have computed  the temperature dependence of the  single particle spectral 
function for the one-dimensional Hubbard model, for a parameter range which has 
been proposed \cite{Claessen03} for the  modeling of the TCNQ band in  TTF-TCNQ organics; 
$ U/t = 4.9  $, $ t = 0.4 eV$  and  $n = 0.59$.  This parameter set reproduces well  the
overall features of the  photoemission spectra at $T=60K$ above the Peierls temperature.  
For this parameter set we have 
identified a magnetic energy scale $T_J \simeq 0.1 t $ below  which $2k_f $ spin fluctuations
are enhanced as a function of decreasing temperature.  For $ T < T_J $ the overall features 
of the $T=0$  spectral function are apparent. In particular no shift in spectral weight between 
holon, holon-shadow and spinon branches is observed below this temperature scale. On the 
other hand,  for $T > T_J$ spectral weight transfer over an energy scale set by $t$ is observed. 
Very similar conclusions  have been reached  for the half-filled Hubbard model \cite{Bulut05}. 
With  $t = 0.4eV $ we obtain a magnetic scale $T_J \simeq 400 K$ and hence we are unable to 
account for the   spectral weight transfer observed in the photoemission experiments 
in the temperature range  $60 K < T < 260 K $ \cite{Claessen03}.  

Assuming that a pure electronic model is valid to account for the temperature 
dependence of the spectral function, other parameter sets are required to understand the 
experimental data. The aim is to keep the low temperature spectral function similar to that observed 
in this work since it compares well with the low temperature photoemission data, 
but to reduce the spin scale $T_J$.  X-ray scattering experiments of 
Ref. \cite{Pouget76}  suggest that both $4k_f$ and to $2k_f$ charge fluctuations are 
present at low temperatures and that above $150 K$ only $4k_f$ scattering is present. 
To model such dominant $4k_f$ fluctuations one requires a Luttinger liquid 
parameter  $ K_\rho < 1/2$ \cite{Schulz90}.  Since the 
Hubbard model has $ 1/2  < K_\rho  < 1  $ additional terms such as a nearest neighbor Coulomb 
repulsion $V$ is  required.  However, we expect that $V$-terms in the Hamiltonian will 
enhance the  overall low temperature {\it band} width.  This band-width problem could be 
corrected by 
reducing the value of the hopping matrix element to it's bulk value, $ t \sim 0.2 eV$,
as inferred from DFT  calculations \cite{Claessen03}. In turn this would enhance the value of 
$U/t$ and hence reduce the value of $T_J$.  
Further simulations are required to confirm this point of view. 

The issue of coupling to the lattice is still open. In particular since the 
system is close to a Peierls transition, it is not clear that phonons can be omitted. 
Furthermore the line-shapes of  model calculations at low temperature  are much sharper 
than the experimentally observed. Coupling to the lattice, could account for this broadening.

\vspace*{1cm}
\noindent 
{\bf Acknowledgments.}
Part of the calculations presented here were carried out on the IBM p690 cluster of
the NIC  in J\"ulich.  We  would like to thank this institution for  allocation of CPU time.
We have greatly profited from discussions with R. Claessen and would like to thank 
E. Jeckelmann for sending us the DDMRG results.  
Similar issues concerning the temperature behavior of the 
spectral function for TTF-TCNQ  have been addressed by N. Bulut, H. Matsueda, T. Tohoyama 
and S. Maekawa.   We would like to thank those authors for sending us a version of 
their manuscript prior to publication.  Financial support from the 
{\it Centre de Coop\'eration Universitaire Franco-Bavarois} (CCUFB-BFHZ)   is acknowledge.


\begin{thebibliography}{10}

\bibitem{Giamarchi}
T. Giamarchi, {\em Quantum physics in one dimension} (Clarendon Press, Oxford,
  2004), iSBN 0 19 85 25 00 1.

\bibitem{Penc96}
K. Penc, K. Hallberg, F. Mila, and H. Shiba, Phys. Rev. Lett {\bf 77},  1390
  (1996).

\bibitem{Jeckelmann04}
H. Benthien, F. Gebhard, and E. Jeckelmann, Phys. Rev. Lett. {\bf 69},  256401
  (2004).

\bibitem{Preuss94}
R. Preuss, A. Muramatsu, W. von~der Linden, P. Dieterich, F. Assaad, and W.
  Hanke, Phys. Rev. Lett. {\bf 73},  732  (1994).

\bibitem{Claessen03}
M. Sing, U. Schwingenschl\"ogl, R. Claessen, P. Blaha, J.~M.~P. Carmelo, L.~M.
  Martelo, P.~D. Sacramento, M. Dressel, and C.~S. Jacobsen, Phys. Rev. B {\bf
  68},  125111  (2003).

\bibitem{Assaad02}
F.~F. Assaad,  in {\em Lecture notes of the Winter School on Quantum
  Simulations of Complex Many-Body Systems :From Theory to Algorithms.}, edited
  by J. Grotendorst, D. Marx, and A. Muramatsu. (Publication Series of the John
  von Neumann Institute for Computing., ADDRESS, 2002), Vol.~NIC series Vol.
  10., pp.\ 99--155.

\bibitem{Beach04a}
K.~S.~D. Beach, cond-mat/0403055  (2004).

\bibitem{Jarrell96}
M. Jarrell and J. Gubernatis, Physics Reports {\bf 269},  133  (1996).

\bibitem{Linden95}
W. von~der Linden, Applied Physics A {\bf 60},  155  (1995).

\bibitem{Schulz90}
H. Schulz, Phys. Rev. Lett {\bf 64},  2831  (1990).

\bibitem{Schulz95}
H. Schulz, Cond-mat/9503150  .

\bibitem{Bulut05}
H. Matsueda, N. Bulut, T. Tohyama, and S. Maekawa, Phys. Rev. B {\bf 72},
  075136  (2005).

\bibitem{Pouget76}
J. Pouget, S.~K. Khanna, F. Denoyer, R. Com\`es, A.~F. Garito, and A.~J.
  Heeger, Phys. Rev. Lett. {\bf 37},  437  (1976).

\end{thebibliography}

\end{document}